\documentclass[preprint, 3p, 12pt,authoryear]{elsarticle}

\usepackage{amssymb}
\usepackage{amsthm}
\usepackage{amsmath} 
\usepackage{enumitem}
\usepackage{tikz}
\usetikzlibrary{positioning}
\usepackage{pifont}
\usepackage{array}
\usepackage{float}
\usepackage[colorlinks=true, citecolor=blue, linkcolor=blue, urlcolor=blue]{hyperref}
\usepackage{comment}

\journal{Nuclear Physics B}

\begin{document}

\begin{frontmatter}



\title{Attributing Forecast Gaps to Component Models in Complex Model Suites}


\author[1]{Xuan Mei\fnref{fn1}}
\ead{xuan.mei@chase.com}

\author[1]{Junze Lin\fnref{fn1}}
\ead{junze.lin@jpmorgan.com}

\affiliation[1]{organization={JPMorgan Chase \& Co.},
	addressline={545 Washington Blvd.}, 
	city={Jersey City},
	postcode={07310}, 
	state={NJ},
	country={USA}}

\fntext[fn1]{The authors work in the Wholesale Credit QR group at JPMorgan Chase \& Co. This work was prepared in the authors' personal capacity and interest. The views expressed are solely those of the authors and do not represent those of JPMorgan Chase \& Co.}

\begin{abstract}
Complex model suites composed of multiple interacting component models are widely used in financial forecasting and risk management. In model performance testing, including in-sample backtesting (BT) and out-of-sample ongoing performance monitoring (OPM), a material gap between a model-suite forecast and the realized outcome must often be attributed to individual component models for development, validation, and regulatory review. This paper studies this gap-attribution problem in the expected loss framework, where exposure at default (EAD), prepayment or single monthly mortality (SMM), probability of default (PD), and loss given default (LGD) interact multiplicatively and are aggregated across loans and projection periods. We first formalize standard walk analysis and show why its attribution is generally order dependent. We then adapt two order-independent attribution frameworks: an augmented Logarithmic Mean Divisia Index (LMDI) approach tailored to the expected-loss structure, and a more general Shapley value approach based on averaging marginal contributions over all component orderings. We derive both elementwise and vectorized formulas to support efficient implementation, with the additional computation time for gap attribution typically limited to a few seconds in practical portfolio-scale examples. Finally, we discuss the connections among walk analysis, LMDI, and Shapley attribution, and show how the attribution framework extends to model suites with an additional Monte Carlo simulation layer.

\end{abstract}

\begin{graphicalabstract}
\centering
\begin{tikzpicture}[
    node distance=1.15cm,
    box/.style={draw, rounded corners, align=center, minimum width=4.6cm, minimum height=0.85cm, font=\small},
    method/.style={draw, rounded corners, align=center, minimum width=2.6cm, minimum height=0.8cm, font=\small},
    arrow/.style={->, line width=0.8pt}
]
\node[box] (forecast) {Model-suite forecast EL from \\{\footnotesize EAD, SMM, PD, LGD}};
\node[box, below=0.85cm of forecast] (gap) {Observed forecast gap\\{\footnotesize forecast vs. actual outcome}};

\node[method, below left=1.05cm and 3.15cm of gap] (walk) {Walk\\analysis};
\node[method, below=1.05cm of gap] (lmdi) {Augmented\\LMDI};
\node[method, below right=1.05cm and 3.15cm of gap] (shapley) {Shapley\\value};

\node[box, below=1.3cm of lmdi] (attrib) {Component-level attribution\\{\footnotesize EAD gap, SMM gap, PD gap, LGD gap}};
\node[box, below=0.85cm of attrib] (extension) {Vectorized implementation\\and Monte Carlo extension};

\draw[arrow] (forecast) -- (gap);
\draw[arrow] (gap) -- (walk);
\draw[arrow] (gap) -- (lmdi);
\draw[arrow] (gap) -- (shapley);
\draw[arrow] (walk) -- (attrib);
\draw[arrow] (lmdi) -- (attrib);
\draw[arrow] (shapley) -- (attrib);
\draw[arrow] (attrib) -- (extension);
\end{tikzpicture}
\end{graphicalabstract}

\begin{highlights}
\item Forecast gaps in expected-loss model suites are attributed to component models.
\item Walk analysis is formalized and its order dependence is explained.
\item Augmented LMDI and Shapley-value methods provide order-independent attribution.
\item Elementwise and vectorized formulas support fast portfolio-scale implementation.
\item The framework extends naturally to model suites with Monte Carlo simulation layers.
\end{highlights}


\end{frontmatter}


\section{Problem Setup}
\subsection{Expected Loss Forecasting}

In credit risk, complex model suites are commonly used to capture portfolio dynamics and produce forecasts. As an example, consider the expected loss (EL) framework widely used in banking; see, for example, \cite{fed-credit-risk-models}. For loan $i$ in period $t$ of the projection horizon, where a period can be a month or a quarter, expected loss is calculated as
\begin{equation}\label{eq11}
\mathrm{EL}_{i,t} = \mathrm{EAD}_{i,t}\times \mathrm{PD}_{i,t}\times \mathrm{LGD}_{i,t}
\end{equation}
where $\mathrm{EAD}_{i,t}$ is exposure at default, $\mathrm{PD}_{i,t}$ is probability of default, and $\mathrm{LGD}_{i,t}$ is loss given default. The EAD component may itself depend on prepayment and utilization models. The portfolio-level expected loss over the projection horizon is then the sum over all $N$ loans and $T$ periods:
\begin{equation}\label{eq12}
\mathrm{EL} = \sum_{i=1}^{N}\sum_{t=1}^{T} \mathrm{EL}_{i,t}
\end{equation}

Model suites are evaluated through both in-sample tests, such as backtesting (BT), and out-of-sample tests, such as ongoing performance monitoring (OPM). In these exercises, the final model-suite forecast, such as portfolio EL, is compared with realized outcomes. When a material gap is observed, model developers, validators, and regulators often need to understand how much of the gap is attributable to each component model.

\subsection{Attribution Challenges}
Several features make this attribution problem nontrivial.
\begin{enumerate}[label=\arabic*)]
\item Component models are coupled in a non-additive way. For example, in equation \eqref{eq11}, loan-level EAD, PD, and LGD are combined multiplicatively and then aggregated across loans and periods in equation \eqref{eq12}. The resulting forecast cannot be decomposed by a simple additive identity.
\item Component-model outputs are often correlated. For example, loans with high loan-to-value (LTV) ratios may have lower prepayment rates, higher PDs, and higher LGDs. Similarly, financially stressed borrowers may be more likely to draw unused credit lines, affecting both EAD and LGD.
\item Realized outcomes for component models are often dependent or partially observed. For example, default and prepayment are competing events, while realized LGD is observed only after default and is not observed for non-defaulted loans.
\end{enumerate}

\subsection{Some Notes}
Three notes are worth emphasizing.

First, to make the paper useful to a broader range of readers while protecting confidential company information, we keep the discussion model-agnostic. We assume only that the model suite follows the general credit-loss framework outlined in \eqref{eq11} and \eqref{eq12}. The individual component models may range from simple linear regressions to recurrent neural networks, and their features may include transformations of loan attributes, demographic variables, and macroeconomic variables such as unemployment rates and market indices. The specification of these models is not the focus of this paper.

Second, in some credit-risk domains, the individual component models are embedded within a Monte Carlo simulation layer with $M$ independent paths: path-level expected loss is first calculated as specified in \eqref{eq11} and \eqref{eq12} in every path, and then averaged over all $M$ paths. In that case, all quantities in \eqref{eq11} carry an additional subscript $m$, denoting the $m$th Monte Carlo path, and \eqref{eq12} becomes
\begin{equation}\label{eq13}
\mathrm{EL} = \frac{1}{M}\sum_{m=1}^M\sum_{i=1}^{N}\sum_{t=1}^{T} \mathrm{EL}_{i,t,m}
\end{equation}

Lastly, while we focus on the gap between model forecasts and actual outcomes, methodologies we discuss in this paper can be seamlessly applied to the gap of forecasts from two versions of model suites.  

\subsection{Structure of This Paper}
The rest of the paper is structured as follows.

Because this is a specialized topic with limited published research, we do not include a separate literature review section. Instead, we introduce the relevant decomposition ideas directly in the methodological sections where they are used.

Section~\ref{Section2} presents the main attribution methods. Section~\ref{subsection_walk_analysis} formalizes standard walk analysis, derives its attribution formulas, and explains the source of its order dependence. Section~\ref{section_LMDI} adapts the Logarithmic Mean Divisia Index (LMDI) to the expected-loss setting, including the special treatment needed for corner cases such as zero values and partially observed LGD outcomes. Section~\ref{subsection_shapley} develops the Shapley decomposition, first through the characteristic-function setup and elementwise formulas, and then through vectorized formulas for efficient implementation and fast computation. Section~\ref{subsection_connection} discusses the connections among walk analysis, LMDI, and Shapley attribution, including the interpretation of Shapley attribution as an average over all walk orders and the relationship between LMDI and Shapley attribution under a log-transformed expected-loss target.

Section~\ref{Section3} extends the attribution framework to model suites with an additional Monte Carlo simulation layer. It shows how path-level attributions can be averaged, or more generally linearly combined with path weights, to obtain the final component-level attribution.

\section{Attribution Methods}\label{Section2}
\subsection{Walk Analysis}\label{subsection_walk_analysis}
A common practical approach is walk analysis. In a ``Prepayment $\to$ PD $\to$ LGD'' walk, one starts with the model-predicted loss and then sequentially replaces model-predicted components with their realized counterparts. The change at each step is assigned to the component replaced at that step. This produces an intuitive attribution from the model forecast to the realized outcome. However, because the components interact multiplicatively and may be correlated, the resulting attribution can depend materially on the order of the walk.

\subsubsection{A Simple Order-Dependence Example}

As a simple demonstration, consider the multiplicative model Sales $=$ Price $\times$ Volume. Actual Price, Volume, and Sales are \$1, 10, and \$10, while the model predicts \$2, 16, and \$32, respectively, as summarized in Table \ref{tab:example-price-volume}. The total sales gap is \$32 - \$10 = \$22. Table \ref{tab:price_attribution} reports the attribution under two walk orders: ``Price $\to$ Volume'', which replaces Price first and Volume second, and ``Volume $\to$ Price'', which reverses the order.


\begin{table}[H]
\centering
\caption{A Simple Price and Volume Example}
\label{tab:example-price-volume}
\begin{tabular}{>{\centering\arraybackslash}p{3cm}|>{\centering\arraybackslash}p{3.8cm} >{\centering\arraybackslash}p{3.8cm} >{\centering\arraybackslash}p{3.8cm}}
\hline
 & Price per Unit (\$)  & Volume (Units Sold) & Total Sales (\$)\\
\hline
Actual  & 1 & 10 & 10 \\

Predicted  & 2 & 16 & 32 \\
\hline
\end{tabular}
\end{table}


\begin{table}[H]
\centering
\caption{Gap Attribution with Different Walk Orders}
\label{tab:price_attribution}
\begin{tabular}{>{\centering\arraybackslash}p{3cm}|>{\centering\arraybackslash}p{3.8cm} >{\centering\arraybackslash}p{3.8cm} >{\centering\arraybackslash}p{3.8cm}}
\hline
 Walk Order & Price   &  Volume & Total Gap \\
\hline
Price       $\to$ Volume  & 16 & 6 & 22 \\
Volume  $\to$ Price  & 10 & 12 & 22 \\
\hline
\end{tabular}
\end{table}
This example shows that the walk order matters. The ``Price $\to$ Volume'' walk suggests that Price contributes more to the gap (16 from Price versus 6 from Volume), whereas the ``Volume $\to$ Price'' walk suggests that Volume contributes more (12 from Volume versus 10 from Price).

\subsubsection{Mathematical Source of Order Dependence}
To see the issue more clearly, denote actual Price and Volume by $P$ and $V$, respectively, and model-predicted Price and Volume by $P+\Delta P$ and $V+\Delta V$. Table \ref{tab:price_attribution_math} gives the corresponding mathematical decomposition.
\begin{figure}[H]
\caption{Diagram of Walk Analyses with Different Orders, Mathematical Explanation}
\centering
\begin{tikzpicture}[>=stealth, line width=0.9pt]
  \node (start) at (0,0) {$(P+\Delta{P})\times (V+\Delta{V})$};
  \coordinate (topL) at (4.92,1.8);
  \coordinate (topR) at (5.08,1.8);
  \node[above=3pt] at (5,1.8) {$P\times(V+\Delta{V})$};
  \coordinate (bottomL) at (4.92,-1.8);
  \coordinate (bottomR) at (5.08,-1.8);
  \node[below=3pt] at (5,-1.8) {$(P+\Delta{P})\times V$};
  \node (end) at (10,0) {$P\times V$};

  \draw[->] (start) -- (topL) 
    node[midway, above, sloped, font=\footnotesize] {plug in actual  $P$};
  \draw[->] (topR) -- (end) 
    node[midway, above, sloped, font=\footnotesize] {plug in actual  $V$ };
  \draw[->] (start) -- (bottomL) 
    node[midway, below, sloped, font=\footnotesize] {plug in actual $V$};
  \draw[->] (bottomR) -- (end) 
    node[midway, below, sloped, font=\footnotesize] {plug in actual $P$};
\end{tikzpicture}

\label{fig:gap-attribution-paths-2}
\end{figure}
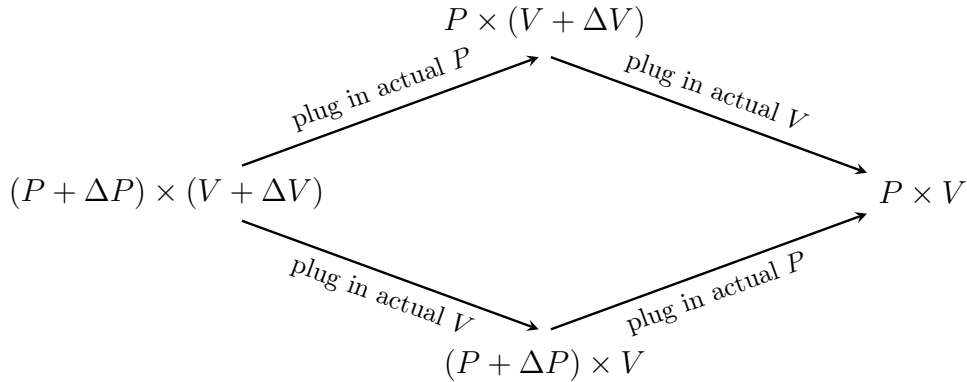

\begin{table}[htbp]
\centering
\caption{Gap Attribution with Different Walk Orders, Mathematical Explanation}
\label{tab:price_attribution_math}
\begin{tabular}{>{\centering\arraybackslash}p{3cm}|>{\centering\arraybackslash}p{3.8cm} >{\centering\arraybackslash}p{3.8cm} }
\hline
 Walk Order & Price   &  Volume \\
\hline
Price       $\to$ Volume  & $\Delta{P}\times V + \Delta{P}\times \Delta{V}  $ & $P\times \Delta{V}$  \\
Volume  $\to$ Price  & $\Delta{P}\times V$ & $P \times \Delta{V} + \Delta{P}\times \Delta{V}$ \\
\hline
\end{tabular}
\end{table}

In walk analysis, the gap between model-estimated and actual sales is
$$\Delta{S} = (P+\Delta{P})\times (V+\Delta{V}) - P \times V = P\times \Delta{V} + \Delta{P}\times V + \Delta{P}\times\Delta{V}$$
The interaction term $\Delta P\times \Delta V$ is entirely assigned to the component whose actual value is plugged in first. When Price and Volume errors are positively correlated, so that $\Delta P$ and $\Delta V$ have the same sign, this interaction term is positive and can unfairly make the first component in the walk appear worse.

\subsubsection{Attribution Formulas}
Because attributions from walk analysis are order dependent, we recommend traversing all possible walk orders and then taking their average, which corresponds to the Shapley decomposition described in Section~\ref{subsection_shapley}.  

Using the notation and formulas developed in Section~\ref{attribution_vector}, attributions under a particular walk can be calculated in less than one second. 

Taking the ``Prepayment$\to$PD$\to$LGD'' walk as an example, we can write
\begin{align*}
\mathrm{EL}^{\mathrm{Model}} - \mathrm{EL}^{\mathrm{Actual}} &= 	\Big(\mathrm{EL}^{\mathrm{Model}} -  f(\mathbf{PD}^{\mathrm{Model}}, \mathbf{LGD}^{\mathrm{Model}}, \mathbf{SMM}^{\mathrm{Actual}}) \Big) \nonumber \\
&+ \Big( f(\mathbf{PD}^{\mathrm{Model}}, \mathbf{LGD}^{\mathrm{Model}}, \mathbf{SMM}^{\mathrm{Actual}}) - f(\mathbf{PD}^{\mathrm{Actual}}, \mathbf{LGD}^{\mathrm{Model}}, \mathbf{SMM}^{\mathrm{Actual}})\Big) \nonumber \\
&+ \Big( f(\mathbf{PD}^{\mathrm{Actual}}, \mathbf{LGD}^{\mathrm{Model}}, \mathbf{SMM}^{\mathrm{Actual}}) - \mathrm{EL}^{\mathrm{Actual}} \Big)
\end{align*}
The three quantities on the right-hand side are the attributions to the prepayment, PD, and LGD models, respectively. 

\subsection{Logarithmic Mean Divisia Index (LMDI)} \label{section_LMDI}
\subsubsection{Formula}
For the multiplicative structure in \eqref{eq11} and \eqref{eq12}, one possible alternative is the Logarithmic Mean Divisia Index (LMDI; see \cite{ang2005lmdi}).

Taking $\ln(\cdot)$ on both sides of \eqref{eq11} gives $$\ln(\mathrm{EL}_{i,t}) = \ln(\mathrm{EAD}_{i,t}) + \ln(\mathrm{PD}_{i,t}) + \ln(\mathrm{LGD}_{i,t}).$$
With this, we get 
\begin{align}
\mathrm{EL}^{\mathrm{Model}}_{i,t} - \mathrm{EL}^{\mathrm{Actual}}_{i,t} & = \displaystyle \frac{\mathrm{EL}^{\mathrm{Model}}_{i,t} - \mathrm{EL}^{\mathrm{Actual}}_{i,t} }{\ln(\mathrm{EL}^{\mathrm{Model}}_{i,t}) - \ln(\mathrm{EL}^{\mathrm{Actual}}_{i,t}) }\cdot \Bigl[(\ln(\mathrm{EAD}^{\mathrm{Model}}_{i,t}) - \ln(\mathrm{EAD}^{\mathrm{Actual}}_{i,t}))\nonumber \\
&+(\ln(\mathrm{PD}^{\mathrm{Model}}_{i,t}) - \ln(\mathrm{PD}^{\mathrm{Actual}}_{i,t})) + (\ln(\mathrm{LGD}^{\mathrm{Model}}_{i,t}) - \ln(\mathrm{LGD}^{\mathrm{Actual}}_{i,t}))\Bigr]\label{lmdi_2}
\end{align}
For term loans without a credit line, let $\mathrm{ScheduleBalance}_{i,t}$ denote the scheduled balance of loan $i$ in period $t$. Given the interest-rate curve, this quantity is deterministic, and

\begin{align*}
\mathrm{EAD}^{\mathrm{Model}}_{i,t} &= \prod_{l=1}^{t-1}(1-\mathrm{PD}^{\mathrm{Model}}_{i,l})(1-\mathrm{SMM}^{\mathrm{Model}}_{i,l})\times \mathrm{ScheduleBalance}_{i,t} \times(1-\mathrm{SMM}^{\mathrm{Model}}_{i, t})\\
\mathrm{EAD}^{\mathrm{Actual}}_{i,t} &= \prod_{l=1}^{t-1}(1-\mathrm{PD}^{\mathrm{Actual}}_{i,l})(1-\mathrm{SMM}^{\mathrm{Actual}}_{i,l})\times \mathrm{ScheduleBalance}_{i,t} \times(1-\mathrm{SMM}^{\mathrm{Actual}}_{i, t})
\end{align*}
where $\mathrm{SMM}^{\mathrm{Model}}_{i,t}$ is the model-estimated prepayment probability for loan $i$ in period $t$, and $\mathrm{SMM}^{\mathrm{Actual}}_{i,t}$ is the realized prepayment indicator. The notation SMM comes from the single monthly mortality rate, but the period can be monthly or quarterly depending on the modeling choice.

With this setup,  $\ln(\mathrm{EAD}^{\mathrm{Model}}_{i,t}) - \ln(\mathrm{EAD}^{\mathrm{Actual}}_{i,t})$ in \eqref{lmdi_2} can be further reduced as
\begin{align*}
& \ln(\mathrm{EAD}^{\mathrm{Model}}_{i,t}) - \ln(\mathrm{EAD}^{\mathrm{Actual}}_{i,t}) \nonumber \\
= &  \sum_{l=1}^{t} (\ln(1-\mathrm{SMM}_{i,l}^{\mathrm{Model}}) - \ln(1-\mathrm{\mathrm{SMM}}_{i,l}^{\mathrm{Actual}}))
+ \sum_{l=1}^{t-1} (\ln(1-\mathrm{PD}_{i,l}^{\mathrm{Model}}) - \ln(1-\mathrm{\mathrm{PD}}_{i,l}^{\mathrm{Actual}}))
\end{align*} 

Equation \eqref{lmdi_2} can then be decomposed as
\begin{align}
& \mathrm{EL}^{\mathrm{Model}}_{i,t} - \mathrm{EL}^{\mathrm{Actual}}_{i,t} \nonumber \\
	=& w_{i,t}\cdot \sum_{l=1}^{t} (\ln(1-\mathrm{SMM}_{i,l}^{\mathrm{Model}}) - \ln(1-\mathrm{\mathrm{SMM}}_{i,l}^{\mathrm{Actual}}))  \nonumber \\
& + w_{i,t}\cdot \Bigl(\sum_{l=1}^{t-1} (\ln(1-\mathrm{PD}_{i,l}^{\mathrm{Model}}) - \ln(1-\mathrm{\mathrm{PD}}_{i,l}^{\mathrm{Actual}}))+(\ln(\mathrm{PD}^{\mathrm{Model}}_{i,t}) - \ln(\mathrm{PD}^{\mathrm{Actual}}_{i,t})) \Bigr) \nonumber \\
& + w_{i,t}\cdot (\ln(\mathrm{LGD}^{\mathrm{Model}}_{i,t}) - \ln(\mathrm{LGD}^{\mathrm{Actual}}_{i,t}))
\end{align}
Summing this decomposition over all loans and periods gives
\begin{align}
& \mathrm{EL}^{\mathrm{Model}} - \mathrm{EL}^{\mathrm{Actual}} = \sum_i \sum_t \left(\mathrm{EL}^{\mathrm{Model}}_{i,t} - \mathrm{EL}^{\mathrm{Actual}}_{i,t}\right) \nonumber \\
=& \sum_i \sum_t \Bigl( w_{i,t}\cdot \sum_{l=1}^{t} (\ln(1-\mathrm{SMM}_{i,l}^{\mathrm{Model}}) - \ln(1-\mathrm{SMM}_{i,l}^{\mathrm{Actual}})) \Bigr)  \nonumber \\
& + \sum_i \sum_t \Bigl( w_{i,t}\cdot \Bigl(\sum_{l=1}^{t-1} (\ln(1-\mathrm{PD}_{i,l}^{\mathrm{Model}}) - \ln(1-\mathrm{PD}_{i,l}^{\mathrm{Actual}}))+(\ln(\mathrm{PD}^{\mathrm{Model}}_{i,t}) - \ln(\mathrm{PD}^{\mathrm{Actual}}_{i,t})) \Bigr) \Bigr) \nonumber \\
& + \sum_i \sum_t \Bigl( w_{i,t}\cdot (\ln(\mathrm{LGD}^{\mathrm{Model}}_{i,t}) - \ln(\mathrm{LGD}^{\mathrm{Actual}}_{i,t}))\Bigr)  \label{lmdi_decomp_sum}
\end{align}
where $$w_{i,t} = \frac{\mathrm{EL}^{\mathrm{Model}}_{i,t} - \mathrm{EL}^{\mathrm{Actual}}_{i,t} }{\ln(\mathrm{EL}^{\mathrm{Model}}_{i,t}) - \ln(\mathrm{EL}^{\mathrm{Actual}}_{i,t}) }.$$
The three terms on the right-hand side of \eqref{lmdi_decomp_sum} are the attributions to the prepayment, PD, and LGD models, respectively.

\subsubsection{Corner Cases and Special Handling}
The model configuration ensures $\mathrm{SMM}_{i,t}^{\mathrm{Model}}\in(0,1)$, $\mathrm{PD}^{\mathrm{Model}}_{i,t}\in(0,1)$, and $\mathrm{LGD}^{\mathrm{Model}}_{i,t}\in (0,1]$, so the logarithmic terms in \eqref{lmdi_decomp_sum} are well-defined for model outputs.

Because $\mathrm{EL}^{\mathrm{Actual}}_{i,t} \ge 0$ while $\mathrm{EL}^{\mathrm{Model}}_{i,t} >0$, and each component model is calculated to six or even nine decimal places, it is very rare, if ever possible, to observe $\mathrm{EL}^{\mathrm{Model}}_{i,t} - \mathrm{EL}^{\mathrm{Actual}}_{i,t} = 0$.

For realized outcomes, however, the indicators $\mathrm{SMM}_{i,t}^{\mathrm{Actual}}$ and $\mathrm{PD}^{\mathrm{Actual}}_{i,t}$ can frequently be zero or one, and $\mathrm{LGD}^{\mathrm{Actual}}_{i,t}$ can be zero. To avoid $\ln(0)$, we follow the small-$\epsilon$ adjustment recommended in \cite{ang2015lmdi}. In addition, because actual LGD is not observed for non-defaulted loans, we assign no LGD attribution in those cases by setting actual LGD equal to model LGD. These considerations lead to the following adjustment:
\begin{align}
\widetilde{\mathrm{SMM}}^{\mathrm{Actual}}_{i,t}
&=
\begin{cases}
0, & \text{if } \mathrm{SMM}^{\mathrm{Actual}}_{i,t}=0, \\
1-\epsilon, & \text{if } \mathrm{SMM}^{\mathrm{Actual}}_{i,t}=1,
\end{cases} \nonumber \\
\widetilde{\mathrm{PD}}^{\mathrm{Actual}}_{i,t}
&=
\begin{cases}
\epsilon, & \text{if } \mathrm{PD}^{\mathrm{Actual}}_{i,t}=0, \\
1-\epsilon, & \text{if } \mathrm{PD}^{\mathrm{Actual}}_{i,t}=1,
\end{cases} \nonumber \\
\widetilde{\mathrm{LGD}}^{\mathrm{Actual}}_{i,t}
&=
\begin{cases}
\mathrm{LGD}^{\mathrm{Model}}_{i,t}, & \text{if } \mathrm{PD}^{\mathrm{Actual}}_{i,t} = 0  , \\
\epsilon, & \text{if } \mathrm{PD}^{\mathrm{Actual}}_{i,t}=1 \text{ and } \mathrm{LGD}^{\mathrm{Actual}}_{i,t}=0,\\
\mathrm{LGD}^{\mathrm{Actual}}_{i,t},& \text{if } \mathrm{PD}^{\mathrm{Actual}}_{i,t}=1 \text{ and } \mathrm{LGD}^{\mathrm{Actual}}_{i,t}> 0
\end{cases}\label{eq:epsilon-adjustment}
\end{align}
where $\epsilon>0$ is a sufficiently small constant. Several values of $\epsilon$ can be tested to assess the sensitivity of the decomposition. 
Note that
\begin{align*}
\widetilde{\mathrm{EL}}^{\mathrm{Actual}} &= \sum_{i,t}\widetilde{\mathrm{EL}}^{\mathrm{Actual}}_{i,t} =\sum_{i,t} \widetilde{\mathrm{EAD}}^{\mathrm{Actual}}_{i,t}\times \widetilde{\mathrm{PD}}^{\mathrm{Actual}}_{i,t}\times \widetilde{\mathrm{LGD}}^{\mathrm{Actual}}_{i,t} \\
   &= \mathrm{EL}^{\mathrm{Actual}} + O(N\cdot T \cdot B \cdot \epsilon)
\end{align*}
where $N$ is the number of loans, $T$ is the number of periods, and $B = \max_{i,t}(\mathrm{ScheduleBalance}_{i,t})$.

For example, if $N = 40{,}000$, $T = 24$ months, and $B = 10^8$ dollars, setting $\epsilon = 10^{-15}$ keeps $\widetilde{\mathrm{EL}}^{\mathrm{Actual}}$ within approximately one dollar of $\mathrm{EL}^{\mathrm{Actual}}$. A smaller value, such as $\epsilon = 10^{-20}$, can be used as a sensitivity check.

\subsection{Shapley Decomposition}\label{subsection_shapley}
\subsubsection{Definition and Setup}
The Shapley value, originally developed in cooperative game theory, allocates a total value among multiple contributors by averaging each contributor's marginal contribution over all possible orderings. In this setting, each component model receives its average contribution to the gap between model estimates and actual outcomes, making the attribution independent of any arbitrary walk order. The Shapley value is also attractive because it is characterized by efficiency, symmetry, the dummy-player property, and linearity \citep{shapley1953value}.

To define the characteristic function for this problem, use actual EL as the baseline. The three players are the PD, LGD, and prepayment components. A component ``plays'' when its actual value is replaced by the corresponding model estimate in the EL calculation. For example, when only PD plays, the characteristic value is computed using model PD together with actual LGD and actual prepayment:
\begin{equation*}
	v(\mathrm{PD}^{\mathrm{Model}} ) = \mathrm{EL}(\mathrm{PD}^{\mathrm{Model}}, \mathrm{LGD}^{\mathrm{Actual}}, \mathrm{SMM}^{\mathrm{Actual}}) - \mathrm{EL}^{\mathrm{Actual}}
\end{equation*}
This setup naturally satisfies $v(\emptyset) = 0$. 

\subsubsection{Special Treatment for LGD}
The only subtlety is that $\mathrm{LGD}^{\mathrm{Actual}}_{i,t}$ is observed only when $\mathrm{PD}^{\mathrm{Actual}}_{i,t}=1$. To handle non-defaulted loans, define

\begin{align*}
\mathrm{LGD}^{\mathrm{Actual}}_{i,t} = 
\begin{cases}
\mathrm{LGD}^{\mathrm{Actual}}_{i,t}, & \text{if } \mathrm{PD}^{\mathrm{Actual}}_{i,t}=1, \\
\mathrm{LGD}^{\mathrm{Model}}_{i,t}, & \text{if } \mathrm{PD}^{\mathrm{Actual}}_{i,t}=0.
\end{cases}
\end{align*}

This convention does not change $\mathrm{EL}^{\mathrm{Actual}}$.
\subsubsection{Attribution Formulas, Elementwise}
The resulting characteristic values for the $2^3=8$ coalitions are:
\begingroup
\scriptsize
\begin{align*}
v(\emptyset) &= 0, \\
v(\mathrm{PD}^{\mathrm{Model}})
&=
\sum_i \sum_t
\prod_{l=1}^{t-1}
\left(1-\mathrm{PD}^{\mathrm{Model}}_{i,l}\right)
\left(1-\mathrm{SMM}^{\mathrm{Actual}}_{i,l}\right)
\times \mathrm{SB}_{i,t}
\times \left(1-\mathrm{SMM}^{\mathrm{Actual}}_{i,t}\right)
\cdot \mathrm{PD}^{\mathrm{Model}}_{i,t}
\cdot \mathrm{LGD}^{\mathrm{Actual}}_{i,t}
-
\mathrm{EL}^{\mathrm{Actual}},
\\[1em]
v(\mathrm{LGD}^{\mathrm{Model}})
&=
\sum_i \sum_t
\prod_{l=1}^{t-1}
\left(1-\mathrm{PD}^{\mathrm{Actual}}_{i,l}\right)
\left(1-\mathrm{SMM}^{\mathrm{Actual}}_{i,l}\right)
\times \mathrm{SB}_{i,t}
\times \left(1-\mathrm{SMM}^{\mathrm{Actual}}_{i,t}\right)
\cdot \mathrm{PD}^{\mathrm{Actual}}_{i,t}
\cdot \mathrm{LGD}^{\mathrm{Model}}_{i,t}
-
\mathrm{EL}^{\mathrm{Actual}},
\\[1em]
v(\mathrm{SMM}^{\mathrm{Model}})
&=
\sum_i \sum_t
\prod_{l=1}^{t-1}
\left(1-\mathrm{PD}^{\mathrm{Actual}}_{i,l}\right)
\left(1-\mathrm{SMM}^{\mathrm{Model}}_{i,l}\right)
\times \mathrm{SB}_{i,t}
\times \left(1-\mathrm{SMM}^{\mathrm{Model}}_{i,t}\right)
\cdot \mathrm{PD}^{\mathrm{Actual}}_{i,t}
\cdot \mathrm{LGD}^{\mathrm{Actual}}_{i,t}
-
\mathrm{EL}^{\mathrm{Actual}},
\\[1em]
v(\mathrm{PD}^{\mathrm{Model}},\mathrm{LGD}^{\mathrm{Model}})
&=
\sum_i \sum_t
\prod_{l=1}^{t-1}
\left(1-\mathrm{PD}^{\mathrm{Model}}_{i,l}\right)
\left(1-\mathrm{SMM}^{\mathrm{Actual}}_{i,l}\right)
\times \mathrm{SB}_{i,t}
\times \left(1-\mathrm{SMM}^{\mathrm{Actual}}_{i,t}\right)
\cdot \mathrm{PD}^{\mathrm{Model}}_{i,t}
\cdot \mathrm{LGD}^{\mathrm{Model}}_{i,t}
-
\mathrm{EL}^{\mathrm{Actual}},
\\[1em]
v(\mathrm{PD}^{\mathrm{Model}},\mathrm{SMM}^{\mathrm{Model}})
&=
\sum_i \sum_t
\prod_{l=1}^{t-1}
\left(1-\mathrm{PD}^{\mathrm{Model}}_{i,l}\right)
\left(1-\mathrm{SMM}^{\mathrm{Model}}_{i,l}\right)
\times \mathrm{SB}_{i,t}
\times \left(1-\mathrm{SMM}^{\mathrm{Model}}_{i,t}\right)
\cdot \mathrm{PD}^{\mathrm{Model}}_{i,t}
\cdot \mathrm{LGD}^{\mathrm{Actual}}_{i,t}
-
\mathrm{EL}^{\mathrm{Actual}},
\\[1em]
v(\mathrm{LGD}^{\mathrm{Model}},\mathrm{SMM}^{\mathrm{Model}})
&=
\sum_i \sum_t
\prod_{l=1}^{t-1}
\left(1-\mathrm{PD}^{\mathrm{Actual}}_{i,l}\right)
\left(1-\mathrm{SMM}^{\mathrm{Model}}_{i,l}\right)
\times \mathrm{SB}_{i,t}
\times \left(1-\mathrm{SMM}^{\mathrm{Model}}_{i,t}\right)
\cdot \mathrm{PD}^{\mathrm{Actual}}_{i,t}
\cdot \mathrm{LGD}^{\mathrm{Model}}_{i,t}
-
\mathrm{EL}^{\mathrm{Actual}},
\end{align*}
\endgroup
\begin{equation}\label{eq_char_formula}
v(\mathrm{PD}^{\mathrm{Model}}, \mathrm{LGD}^{\mathrm{Model}},\mathrm{SMM}^{\mathrm{Model}}) = \mathrm{EL}^{\mathrm{Model}} - \mathrm{EL}^{\mathrm{Actual}}
\end{equation}

The Shapley decomposition gives
$$\mathrm{EL}^{\mathrm{Model}} - \mathrm{EL}^{\mathrm{Actual}} = v(\mathrm{PD}^{\mathrm{Model}}, \mathrm{LGD}^{\mathrm{Model}},\mathrm{SMM}^{\mathrm{Model}}) = \phi(\mathrm{PD}^{\mathrm{Model}}) + \phi(\mathrm{LGD}^{\mathrm{Model}})+ \phi(\mathrm{SMM}^{\mathrm{Model}}) $$
The attributions to the PD, LGD, and prepayment models are therefore calculated as follows, with the zero-valued $v(\emptyset)$ term omitted:

\begin{equation}
\resizebox{0.88\textwidth}{!}{$\begin{aligned}
\phi(\mathrm{PD}^{\mathrm{Model}})
&= \frac{1}{3}\cdot v(\mathrm{PD}^{\mathrm{Model}})\\
&\quad + \frac{1}{6}\cdot\left[v(\mathrm{PD}^{\mathrm{Model}},\mathrm{LGD}^{\mathrm{Model}})-v(\mathrm{LGD}^{\mathrm{Model}})
+ v(\mathrm{PD}^{\mathrm{Model}},\mathrm{SMM}^{\mathrm{Model}})-v(\mathrm{SMM}^{\mathrm{Model}})\right]\\
&\quad + \frac{1}{3}\cdot\left[v(\mathrm{PD}^{\mathrm{Model}},\mathrm{LGD}^{\mathrm{Model}},\mathrm{SMM}^{\mathrm{Model}})
- v(\mathrm{LGD}^{\mathrm{Model}},\mathrm{SMM}^{\mathrm{Model}})\right],\\
\phi(\mathrm{LGD}^{\mathrm{Model}})
&= \frac{1}{3}\cdot v(\mathrm{LGD}^{\mathrm{Model}})\\
&\quad + \frac{1}{6}\cdot\left[v(\mathrm{PD}^{\mathrm{Model}},\mathrm{LGD}^{\mathrm{Model}})-v(\mathrm{PD}^{\mathrm{Model}})
+ v(\mathrm{LGD}^{\mathrm{Model}},\mathrm{SMM}^{\mathrm{Model}})-v(\mathrm{SMM}^{\mathrm{Model}})\right]\\
&\quad + \frac{1}{3}\cdot\left[v(\mathrm{PD}^{\mathrm{Model}},\mathrm{LGD}^{\mathrm{Model}},\mathrm{SMM}^{\mathrm{Model}})
- v(\mathrm{PD}^{\mathrm{Model}},\mathrm{SMM}^{\mathrm{Model}})\right],\\
\phi(\mathrm{SMM}^{\mathrm{Model}})
&= \frac{1}{3}\cdot v(\mathrm{SMM}^{\mathrm{Model}})\\
&\quad + \frac{1}{6}\cdot\left[v(\mathrm{PD}^{\mathrm{Model}},\mathrm{SMM}^{\mathrm{Model}})-v(\mathrm{PD}^{\mathrm{Model}})
+ v(\mathrm{LGD}^{\mathrm{Model}},\mathrm{SMM}^{\mathrm{Model}})-v(\mathrm{LGD}^{\mathrm{Model}})\right]\\
&\quad + \frac{1}{3}\cdot\left[v(\mathrm{PD}^{\mathrm{Model}},\mathrm{LGD}^{\mathrm{Model}},\mathrm{SMM}^{\mathrm{Model}})
- v(\mathrm{PD}^{\mathrm{Model}},\mathrm{LGD}^{\mathrm{Model}})\right]. 
\end{aligned}$}
\label{eq_shap_sol}
\end{equation}

\subsubsection{Attribution Formulas, Vectorized}\label{attribution_vector}
For efficient implementation and fast computation, the elementwise formulas \eqref{eq_char_formula} and \eqref{eq_shap_sol} can be written in matrix form.
Let $\mathbf{PD}$, $\mathbf{LGD}$, $\mathbf{SMM}$, and $\mathbf{SB}$ be $N\times T$ dimensional matrices, with rows representing loans and columns representing periods. Define the function $f$ on these matrices as
\begin{align}
	f(\mathbf{PD}, \mathbf{LGD}, \mathbf{SMM} ) = & \mathrm{np.sum}\Big(\mathrm{np.cumprod}(1- \big[0_n, \mathbf{PD}[:, :-1]\big], \mathrm{axis} = 1) \nonumber \\
	&* \mathrm{np.cumprod}(1-\mathbf{SMM}, \mathrm{axis} = 1) * \mathbf{SB} * \mathbf{PD} *\mathbf{LGD} \Big) \label{matrix_formula}
\end{align}

Then
$$\mathrm{EL}^{\mathrm{Model}}  = f(\mathbf{PD}^{\mathrm{Model}}, \mathbf{LGD}^{\mathrm{Model}}, \mathbf{SMM}^{\mathrm{Model}} )$$
and
$$ \mathrm{EL}^{\mathrm{Actual}} = f(\mathbf{PD}^{\mathrm{Actual}}, \mathbf{LGD}^{\mathrm{Actual}}, \mathbf{SMM}^{\mathrm{Actual}} ).$$
The Shapley attribution in \eqref{eq_shap_sol} can then be evaluated in vectorized form as

\begin{equation}
\resizebox{0.88\textwidth}{!}{$\begin{aligned}
\phi(\mathrm{PD}^{\mathrm{Model}})
&= \frac{1}{3}\cdot \left[f(\mathbf{PD}^{\mathrm{Model}}, \mathbf{LGD}^{\mathrm{Actual}}, \mathbf{SMM}^{\mathrm{Actual}})- \mathrm{EL}^{\mathrm{Actual}}\right]\\
&\quad + \frac{1}{6}\cdot\left[f(\mathbf{PD}^{\mathrm{Model}}, \mathbf{LGD}^{\mathrm{Model}}, \mathbf{SMM}^{\mathrm{Actual}})-f(\mathbf{PD}^{\mathrm{Actual}}, \mathbf{LGD}^{\mathrm{Model}}, \mathbf{SMM}^{\mathrm{Actual}})\right] \\ 
&\quad + \frac{1}{6}\cdot\left[f(\mathbf{PD}^{\mathrm{Model}}, \mathbf{LGD}^{\mathrm{Actual}}, \mathbf{SMM}^{\mathrm{Model}}) - f(\mathbf{PD}^{\mathrm{Actual}}, \mathbf{LGD}^{\mathrm{Actual}}, \mathbf{SMM}^{\mathrm{Model}})\right]\\
&\quad + \frac{1}{3}\cdot\left[\mathrm{EL}^{\mathrm{Model}}
- f(\mathbf{PD}^{\mathrm{Actual}}, \mathbf{LGD}^{\mathrm{Model}}, \mathbf{SMM}^{\mathrm{Model}})\right],\\
\phi(\mathrm{LGD}^{\mathrm{Model}})
&= \frac{1}{3}\cdot \left[f(\mathbf{PD}^{\mathrm{Actual}}, \mathbf{LGD}^{\mathrm{Model}}, \mathbf{SMM}^{\mathrm{Actual}})- \mathrm{EL}^{\mathrm{Actual}}\right]\\
&\quad + \frac{1}{6}\cdot\left[f(\mathbf{PD}^{\mathrm{Model}}, \mathbf{LGD}^{\mathrm{Model}}, \mathbf{SMM}^{\mathrm{Actual}})-f(\mathbf{PD}^{\mathrm{Model}}, \mathbf{LGD}^{\mathrm{Actual}}, \mathbf{SMM}^{\mathrm{Actual}})\right] \\ 
&\quad + \frac{1}{6}\cdot\left[f(\mathbf{PD}^{\mathrm{Actual}}, \mathbf{LGD}^{\mathrm{Model}}, \mathbf{SMM}^{\mathrm{Model}}) - f(\mathbf{PD}^{\mathrm{Actual}}, \mathbf{LGD}^{\mathrm{Actual}}, \mathbf{SMM}^{\mathrm{Model}})\right]\\
&\quad + \frac{1}{3}\cdot\left[\mathrm{EL}^{\mathrm{Model}}
- f(\mathbf{PD}^{\mathrm{Model}}, \mathbf{LGD}^{\mathrm{Actual}}, \mathbf{SMM}^{\mathrm{Model}})\right],\\
\phi(\mathrm{SMM}^{\mathrm{Model}})
&= \frac{1}{3}\cdot \left[f(\mathbf{PD}^{\mathrm{Actual}}, \mathbf{LGD}^{\mathrm{Actual}}, \mathbf{SMM}^{\mathrm{Model}})- \mathrm{EL}^{\mathrm{Actual}}\right]\\
&\quad + \frac{1}{6}\cdot\left[f(\mathbf{PD}^{\mathrm{Model}}, \mathbf{LGD}^{\mathrm{Actual}}, \mathbf{SMM}^{\mathrm{Model}})-f(\mathbf{PD}^{\mathrm{Model}}, \mathbf{LGD}^{\mathrm{Actual}}, \mathbf{SMM}^{\mathrm{Actual}})\right] \\ 
&\quad + \frac{1}{6}\cdot\left[f(\mathbf{PD}^{\mathrm{Actual}}, \mathbf{LGD}^{\mathrm{Model}}, \mathbf{SMM}^{\mathrm{Model}}) - f(\mathbf{PD}^{\mathrm{Actual}}, \mathbf{LGD}^{\mathrm{Model}}, \mathbf{SMM}^{\mathrm{Actual}})\right]\\
&\quad + \frac{1}{3}\cdot\left[\mathrm{EL}^{\mathrm{Model}}
- f(\mathbf{PD}^{\mathrm{Model}}, \mathbf{LGD}^{\mathrm{Model}}, \mathbf{SMM}^{\mathrm{Actual}})\right]. 
\end{aligned}$}
\label{eq_shap_sol_matrix}
\end{equation}
\subsubsection{Discussion of Computational Complexity}
The classical Shapley value calculation is well known to have exponential complexity in the number of players. As shown in \eqref{eq_shap_sol_matrix}, for the three-player setup considered here, it requires $2^3 = 8$ quantities: $\mathrm{EL}^{\mathrm{Model}}$, $\mathrm{EL}^{\mathrm{Actual}}$, and the six additional $f(\cdot)$ quantities. However, this $2^k$ complexity applies only to the matrix calculation in \eqref{matrix_formula}, which takes no more than 1--2 seconds for $N=40{,}000$ and $T=24$ in our tests. The most computationally intensive quantities, $\mathbf{PD}^{\mathrm{Model}}$, $\mathbf{LGD}^{\mathrm{Model}}$, $\mathbf{SMM}^{\mathrm{Model}}$, and $\mathbf{SB}^{\mathrm{Model}}$, are already derived and stored when calculating $\mathrm{EL}^{\mathrm{Model}}$. Hence, in any BT or OPM exercise where $\mathrm{EL}^{\mathrm{Model}}$ and $\mathrm{EL}^{\mathrm{Actual}}$ are always required and have already been calculated, the additional computational cost of the Shapley attribution solution is negligible.

\subsection{Connections between Different Attributions}\label{subsection_connection}
It is useful to highlight the connections among walk analysis, LMDI, and Shapley attribution. 
\subsubsection{Walk Analysis and Shapley}
The Shapley attribution for a component model is simply the average of its walk attribution across all possible walks. 

Assume there are $k$ component models, collected in the set $K=\{1,\ldots,k\}$. The order of a specific walk is a permutation of these $k$ models, so there are $k!$ possible walk orders. Let $\Pi(K)$ denote the set of all permutations of $K$. For a permutation $\pi\in\Pi(K)$, let $P_i^\pi$ denote the set of models that appear before model $i$ in that permutation. If $v(S)$ denotes the gap obtained after replacing the actual values by model values for the subset $S\subseteq K$, then in the walk $\pi$ the attribution to model $i$ is simply $v\left(P_i^\pi\cup\{i\}\right)-v\left(P_i^\pi\right)$. To make the attribution order-independent, a natural solution is to average across all $k!$ permutations, or equivalently all possible walks:
\begin{equation}\label{eq_shapley_permutation}
\phi_i
= \frac{1}{k!}\sum_{\pi\in\Pi(K)}
\left[v\left(P_i^\pi\cup\{i\}\right)-v\left(P_i^\pi\right)\right]
\end{equation}
Equivalently, it can be written as the coalition-weighted formula
\begin{equation}\label{eq_shapley_subset}
\phi_i
= \sum_{S\subseteq K\setminus\{i\}}
\frac{|S|!(k-|S|-1)!}{k!}
\left[v(S\cup\{i\})-v(S)\right]
\end{equation}

\subsubsection{LMDI and Shapley}
Under the expected loss setup, LMDI can be viewed as a Shapley attribution based on a different characteristic function, followed by a rescaling step. It preserves three of the four properties satisfied by Shapley attribution: efficiency, symmetry, and the null-player property. Although it does not satisfy the standard linearity property in general, it still preserves a special summability property under our expected-loss setup. 

To see this, focus on the $i$th loan in period $t$. Taking $\ln(\cdot)$ on both sides of \eqref{eq11} gives
$$\ln(\mathrm{EL}_{i,t}) = \ln(\mathrm{EAD}_{i,t}) + \ln(\mathrm{PD}_{i,t}) + \ln(\mathrm{LGD}_{i,t}).$$
If the target is $\ln(\mathrm{EL}_{i,t})$, or equivalently if the characteristic function is $v_{i,t}(\cdot) = \ln(\mathrm{EL}_{i,t})$, then the Shapley attributions for the EAD, PD, and LGD models are simply $\ln(\mathrm{EAD}^{\mathrm{Model}}_{i,t}) - \ln(\mathrm{EAD}^{\mathrm{Actual}}_{i,t})$, $\ln(\mathrm{PD}^{\mathrm{Model}}_{i,t}) - \ln(\mathrm{PD}^{\mathrm{Actual}}_{i,t})$, and $\ln(\mathrm{LGD}^{\mathrm{Model}}_{i,t}) - \ln(\mathrm{LGD}^{\mathrm{Actual}}_{i,t})$, respectively. 

However, because our target is $\mathrm{EL}_{i,t}$ rather than $\ln(\mathrm{EL}_{i,t})$, the relevant characteristic function is $u_{i,t}(\cdot)=\exp(v_{i,t}(\cdot))$. As a result, the efficiency property no longer holds for attributions based directly on $v_{i,t}(\cdot)$. LMDI restores efficiency by using the rescaling quantity
$$w_{i,t} =\frac{u_{i,t}(K) - u_{i,t}(\emptyset) }{v_{i,t}(K) - v_{i,t}(\emptyset)} =\frac{\mathrm{EL}^{\mathrm{Model}}_{i,t} - \mathrm{EL}^{\mathrm{Actual}}_{i,t} }{\ln(\mathrm{EL}^{\mathrm{Model}}_{i,t}) - \ln(\mathrm{EL}^{\mathrm{Actual}}_{i,t}) }.$$

Since $w_{i,t}$ depends nonlinearly on the model-vs-actual EL gap, LMDI does not satisfy the linearity property in general setups. However, as shown in \eqref{lmdi_decomp_sum}, under the expected-loss framework in \eqref{eq11} and \eqref{eq12}, the decomposition remains summable across loans $i$ and periods $t$. This property is sufficient for extending the attribution framework to more complex model suites with an additional Monte Carlo simulation layer. 

\section{Extend to Model Suites with Monte Carlo Simulation}\label{Section3}
Some complex model suites include an additional Monte Carlo simulation layer: there can be $M$ independent paths; in each path, \eqref{eq11} and \eqref{eq12} still hold; and the final EL is an average of the path-level ELs, as formulated in \eqref{eq13}. 

Because the final EL is a linear combination of path-level ELs, walk analysis, LMDI, and Shapley attribution can be applied path by path and then aggregated. Let $\phi_i^{(m)}$ be the attribution to component $i$ in the $m$th path. The final attribution to component $i$ is then the average of its attribution across all paths:
$$\phi_i = \frac{1}{M}\sum_{m = 1}^{M}\phi_i^{(m)}$$

This extension is not limited to equally weighted Monte Carlo paths; it also applies to weighted averages. If the equal weight $1/M$ in \eqref{eq13} is replaced by more general weights $w_m$ satisfying $\sum_{m=1}^{M}w_m = 1$, and the final EL becomes
\begin{equation*}
\mathrm{EL} = \sum_{m=1}^M\left(w_m\cdot \sum_{i=1}^{N}\sum_{t=1}^{T} \mathrm{EL}_{i,t,m}\right),
\end{equation*}
then we can calculate the attribution within each Monte Carlo path and linearly combine the path-level attributions to obtain
$$\phi_i = \sum_{m = 1}^{M}w_m\cdot \phi_i^{(m)}.$$
\bibliographystyle{elsarticle-harv} 


\end{document}